\documentstyle[preprint,aps,epsfig]{revtex}
\begin{document}

\tighten
\draft
\preprint{
\vbox{
\hbox{\today}
\hbox{Tashkent}
}}
\newcommand{\ci}[1]{\cite{#1}}
\newcommand{\lab}[1]{\label{#1}}
\newcommand{\re}[1]{(\ref{#1})}
\newcommand{\bfr}{\begin{flushright}}
\newcommand{\bfl}{\begin{flushleft}}
\newcommand{\efl}{\end{flushleft}}
\newcommand{\efr}{\end{flushright}}
\newcommand{\bc}{\begin{center}}
\newcommand{\ec}{\end{center}}
\newcommand{\be}{\begin{equation}}
\newcommand{\ee}{\end{equation}}
\newcommand{\bea}{\begin{eqnarray}}
\newcommand{\eea}{\end{eqnarray}}
\newcommand{\ba}{\begin{array}}
\newcommand{\ea}{\end{array}}
\newcommand{\edc}{\end{document}}
\newcommand{\ul}{\underline}
\newcommand{\ri}{\rightarrow\infty}
\newcommand{\li}{\leftarrow\infty}
\newcommand{\ra}{\rightarrow}
\newcommand{\la}{\leftarrow}
\newcommand{\ds}{\displaystyle}
\newcommand{\dsf}{\displaystyle\frac}
\newcommand{\dt}{\Delta{t}}
\newcommand{\il}{\int\limits}
\newcommand{\pal}{\partial}
\newcommand{\xxx}{{\it{X}}}
\newcommand{\bone}{{\bf 1}}

\title{Complex Diffusion Monte-Carlo method for the
 systems with complex wave function:\\
test by the simulation of 2$D$ electron in uniform magnetic field}

\author{B. Abdullaev$^1$, M. Musakhanov$^1$, A. Nakamura$^{2}$ }
\address{$^1$
 Theoretical  Physics Dept, Tashkent State University,\\
 Tashkent 700174, Uzbekistan\\
e-mail: bah$\_$abd@iaph.silk.org, yousuf@iaph.silk.org}
\address{$^2$
RIISE, Hiroshima University, Japan}

\maketitle

\begin{abstract}
On the base of Diffusion Monte-Carlo method it is developed a new
Complex Diffusion Monte-Carlo (CDMC) method
allowing to simulate the quantum systems
with complex wave function.
There are no approximations on the calculation
of modulus and phase of wave function in contrast to other methods.
We find that the averaged value of any quantity in CDMC
will have no direct contribution from the phase of the distribution
function but only from the phase of the Green function of the diffusion
equation.
This is  most important and crucial point of CDMC.

We are testing CDMC
by the calculation  of the wave function and the ground state energy of
two-dimensional electron placed into the external uniform magnetic field.
There is an excellent agreement
between simulations' results and an analytical ones.

\end{abstract}

\section{Introduction}

Quantum-mechanical
wave function is an essentially  complex in many cases.
There are well-known examples --
electron in the external
uniform magnetic field when vector potential has a central-symmetric
form and system of anyons. The complexity of wave function does not give
a possibility for the simulations of such kind of systems by using
well-known  Green Function
Monte-Carlo method (see review \ci{1}).
This method essentially demand
the reality of the system's wave function which considered
as probability weight in during stochastic process.

There had been undertaken  the several attempts \ci{4,5,6,7}
to construct a Monte-Carlo method for the simulation
of quantum systems with complex wave function.
Authors of the works \ci{4,7} suggested
to take as a probability weight the module of
the complex distribution function.
All quantities are calculated
in \ci{4,7} by averaging over this complex distribution function.
The phase of this function $\alpha $ is taken accordingly
$ \alpha = \sum \alpha_i$, where $\alpha_i $
is the phase at the $i$-time step of the stochastic process.

Another quantum Monte Carlo method using algorithm without
branching for the simulating of
complex problems was developed in \ci{6}.
The main trouble of this approach
was an increasing of the statistical error as a function of
the whole time of the simulation.

  The basic difficulty of the numerical simulation of the fermions
is essentially the same as
for the systems with the complex wave function.
Their wave function can change the sign and therefore
can not be used as the probability weight in the simulation process.

For the simulation of the continuum (not on the lattice) fermionic systems
it were developed widely  used  fixed node Monte-Carlo method
\ci{2}( see also reviews \ci{1,15}) and recently proposed
constrained path Monte-Carlo method \ci{16}. In
these both methods it was assumed the restriction on the random walks
connected with the uncertainties in the space localization of the wave
function node surfaces. The comparison of these methods was done in \ci{17}.

Our method is very close to
fixed phase Diffusion Monte-Carlo method developed in \ci{5},
which was applied
to two-dimensional electrons in magnetic field.
Laughlin's wave function \ci{8} was used as a trial one.
In the framework of Diffusion Monte-Carlo (DMC) method
it was calculated only  modulus of the system's wave function. The phase of
system's wave function was not calculated  but considered as
fixed and equal  to the  phase of Laughlin wave function.

Here it is proposed  new
Complex Diffusion Monte-Carlo (CDMC) method
 including also the simulation of the phase
factor of the wave function.

CDMC are tested first
by the calculations  of the ground state  wave function and energy of
two-dimensional electron placed into the external uniform magnetic field.
 There is an excellent agreement
between simulations' results and an analytical ones.
Namely, it is reproduced the ground state energy
and his degeneracy on the orbital quantum number $m$
and also the simulated wave function
is in exact correspondence with
analytical prediction.
The description of CDMC method is given at the section II of the paper.
In the section III we represent the description of the simulation algorithm.
Analytical expressions for the algorithm quantities are given at section IV.
Section V represent the discussion of the simulation results.

\section{Complex Diffusion Monte-Carlo method for the quantum
systems with complex wave function}

Hamiltonian of the electron placed into external uniform magnetic field
with vector potential $\vec A=\dsf{1}{2}[\vec H\vec r]$, where $\vec H$ -
magnetic field, $\vec r$ - electron's radius vector, and in external potential
$V(r)$ has a form:
\be
\hat H=\dsf{1}{2M}(\vec p+\dsf{|e|}{c}\vec A)^2+ V(r).
\lab{1}
\ee
Here $M$ - mass and $e=-|e|$ charge of electron,  $\vec p=-i\hbar \vec
\nabla$, where  $\vec \nabla=\vec i \dsf{\pal}{\pal x}+
\vec j \dsf{\pal}{\pal y}$, $c$ - light velocity.
By taking in to account a relation  $\vec p\vec A -\vec A\vec p=div \vec A=0$
for the such kind of vector potential one can rewrite Hamiltonian \re{1}
in the following form:
\be
\hat H=\dsf{\vec p\ ^2}{2M}+\dsf{|e|}{Mc}\vec A\vec p+
\dsf{|e|^2}{Mc^2}\vec A^2+ V(r).
\lab{2}
\ee
Let us introduce a complex distribution function
\be
f(\vec r,t)=\Psi_T^*(\vec r)\Psi(\vec r,t).
\lab{3}
\ee
Here $\Psi_T^*(\vec r)$ is
a complex conjugated trial wave function  of electron in magnetic field.
Wave function $\Psi(\vec r,t)$ satisfies a Schrodinger equation with
imaginary time ( expressed in $\hbar$ units )
\be
-\dsf{\pal \Psi(\vec r,t)}{\pal t}=(\hat H-E_T)\Psi(\vec r,t).
\lab{4}
\ee

It is necessary to choose a trial energy $E_T$ in such manner  \ci{2}
that at $t\ri$    $\Psi(\vec r,t)\ra\Psi_0(\vec r)$
- an exact stationary wave function of ground state of Hamiltonian \re{1}.

So, by  accounting \re{4}, one can write the equation for the
distribution function $f\equiv f(\vec r,t)$:
\be
-\dsf{\pal f}{\pal t}=\dsf{\vec p^{\ 2}}{2M}f-
\dsf{\vec p}{M}(f\vec F_Q(\vec r))+\dsf{|e|}{Mc}\vec A\vec p f+
(E_L(\vec r)-E_T)f,
\lab{5}
\ee
where
\be
\vec F_Q(\vec r)=\Psi_T^{*-1}(\vec r)\vec p\Psi_T^{*}(\vec r),
\lab{6}
\ee
\be
E_L(\vec r)=\Psi_T^{*-1}(\vec r)\hat H'\Psi_T^{*}(\vec r),
\lab{7}
\ee
\be
\hat H'=\dsf{1}{2M}(\vec p-\dsf{|e|}{c}\vec A)^2+ V(r).
\lab{8}
\ee

For the $\vec p=-i\hbar\vec \nabla$ and
$D=\dsf{\hbar^2}{2M}$, and by renaming
\be
\vec F_Q(\vec r)=2\Psi_T^{*-1}(\vec r)\vec\nabla\Psi_T^{*}(\vec r),
\lab{9}
\ee
we have an equation:
\be
-\dsf{\pal f}{\pal t}=-D\Delta f+D\vec \nabla(f\vec F_Q(\vec r))-
\dsf{i\hbar|e|}{Mc}\vec A\vec \nabla f+
(E_L(\vec r)-E_T)f.
\lab{10}
\ee
Here $\Delta=\vec \nabla^2$.

In general case  when  $\Psi_T^*(\vec r)$ is a complex wave function
the quantity $\vec F_Q(\vec r)$ has a form:
\be
\vec F_Q(\vec r)=Re \vec F_Q(\vec r)+i Im \vec F_Q(\vec r).
\lab{11}
\ee
So, a final view of an equation for the distribution function
$f$ is
\bea\ba{c}
-\dsf{\pal f}{\pal t}=-D\Delta f+D\vec \nabla(f\ Re \vec F_Q(\vec r))+
i[\vec \nabla(D f\ Im \vec F_Q(\vec r))-\\
-\dsf{\hbar|e|}{Mc}\vec A\vec \nabla f]+
(E_L(\vec r)-E_T)f.
\lab{12}
\ea\eea

Following \ci{2}, we assume: when the time step of integration
of equation \re{12} $\tau \ra 0$  the function $\vec F_Q(\vec r)$ remains
constant, i.e. we will  suppose  that  $\vec F_Q(\vec r)\equiv \vec F_Q(\vec
r\ ')$ at $\tau \ra 0$, where vector $\vec r$ corresponds to time point
$t+\tau$ and $\vec r\ '$ to point $t$.

Let us introduce a new quantity
\be
\vec A_Q(\vec r,\vec r\ ')=\dsf{1}{2}Im \vec F_Q(\vec r\ ')-
\dsf{\hbar|e|}{2DMc} \vec A (\vec r).
\lab{13}
\ee
Then at $\tau \ra 0$ the Green function of equation \re{12} has a form:
\be
G(\vec r,\vec r\ ';\tau)=G_{1}(\vec r,\vec r\ ';\tau)\exp\left[-\tau(E_L(\vec r)-E_T)\right]
\exp\left[i\vec A_Q(\vec r,\vec r\ ')
(\vec r-\vec r\ '-D\tau Re \vec F_Q(\vec r\ '))\right],
\lab{14}
\ee
where
\be
G_{1}(\vec r,\vec r\ ';\tau)=\dsf{\exp[D\tau\vec A_Q^2(\vec r,\vec r\ ')]}{
4\pi D\tau}
\exp\left[-\dsf{(\vec r-\vec r\ '-D\tau Re \vec F_Q(\vec r\ '))^2}{
4D\tau}\right].
\lab{14a}
\ee
One can see that the Green function $G(\vec r,\vec r\ ';\tau)$ \re{14}
and distribution function $f$ given by \re{3} are the complex functions.
These both quantities are related by usual integral equation:
\be
f(\vec r,t+\tau)=\int d\vec r\ ' G(\vec r,\vec r\ ';\tau)f(\vec r\ ',t).
\lab{15}
\ee
From equation \re{15} it is followed that the modulus and the phase of the
distribution function at consequent time point are determined by
the modulus and the phase of the Green function and of ones of the
distribution function at previous time point of integration of
diffusion equation \re{12}.

We have to note that for the general
case of complex wave function
$\Psi_T^*(\vec r)$ the energy $E_L(\vec r)$ is also a complex, so real
and imaginary part of  $E_L(\vec r)$ contribute to ones of the Green function.
Therefore the  last two exponents in \re{14} has a form:
\bea\ba{c}
\exp\left[-\tau(Re E_L(\vec r)-E_T)\right]\times\\
\times\exp\left[i\vec A_Q(\vec r,\vec r\ ')
(\vec r-\vec r\ '-D\tau Re \vec F_Q(\vec r\ '))-i\tau Im E_L(\vec r)\right].
\lab{16}
\ea\eea

\section{ The description of the simulation algorithm}

1) Let us to have the $N_c$ of the initial configurations $\vec r$,
i.e. a set
of initial systems in which the electron position has random
and uniform distribution. One can choose the configurations of $\vec r$ and
with the distribution function $f(\vec r,0)=|\Psi_T(\vec r)|^2$,  because at
$t=0$ \re{3} is real.

The choice of the boundary condition depends on the
problem. For the one electron in the external magnetic field
a boundary is periodic, i.e., if electron
cross a boundary from one side of simulated cell, it enters into cell
from opposite side. For example, for the system of bosons in 2D parabolic
well and
in external magnetic field  the boundaries must be free because parabolic
well determines itself the boundary of space distribution
of the particles.

2) It is performed the quantum drift and diffusion of the particle
from $k$ - th configuration, for example, in accordance with the formula
\be
 \vec r_k=\vec r_k\ '+D\tau Re \vec F_Q(\vec r_k\ ')+\chi.
\lab{17}
\ee

Here  $\chi$ - is a gaussian random number having mean value zero and
dispersion $2\sqrt{D\tau}$.

3) The transition into new space point in this configuration is accepted
with probability
\be
P(\vec r\ '\ra \vec r, \tau)\equiv \min (1,W(\vec r, \vec r\ ')),
\lab{18}
\ee
where
$$W(\vec r, \vec r\ ')=\dsf{|\Psi_T(\vec r)|^2G_1(\vec r\ ', \vec r; \tau)}{
|\Psi_T(\vec r\ ')|^2G_1(\vec r, \vec r\ '; \tau)}.$$
Here
$G_1(\vec r, \vec r\ '; \tau)$
is given by \re{14a}.

If transition of electron is accepted then in accordance with \re{16}
it has  new phase, if no then electron keep his old phase.

4) After changing of the electron position from $k$ - configuration
into new space point,  are calculated $Re E_L(\vec r_k)$, $Im
E_L(\vec r_k)$ and other quantities of interest.

5) By using of the first exponential factor in \re{16} it is calculated the
multiplicity $M_k$ (the branching probability) for the configuration $k$
accordingly
\be
M_k=\exp[-\tau(Re E_L(\vec r_k)-E_T)]
\lab{19}
\ee
If $M_k$ is not integer, we add an uniformly distributed random number
between 0 and 1 to it and take $M_k$ equal to nearest integer.

6) If $M_k\ne 0$ then $M_k$ copies of new $k$-th configuration
place in the list of the new $N$ configurations that one is the initial
at the next step $\tau$ of integration of the diffusion equation.
If $M_k=0$ then there is no $k$-th  configuration in the list of new
$N$ configuration for the next time step $\tau$.

All quantities of interest
as $Re E_L(\vec r_k)$, $Im E_L(\vec r_k)$ and so on
are multiplied by the factor $M_k$ for the calculating of mean values
of these ones.

7) It is repeated steps 2)-6) of algorithm until all
$N_c$ configurations will not overlooked and electrons on these
configurations will not simulated on the displacement and the having a
new phase.

8)  It are calculated a mean energy and
others mean quantities on the $N$ number of configurations, got in
the point 6) of algorithm,
at this time step  $\tau$
in accordance with formula \re{23} (see below, where it is necessary to
change $M$ by $N$ and to take into account that the phase of electron is equal
to the phase of the Green function $\alpha_G$ and at the calculation of a
mean energy  that the energy $E_L(\vec r_k)$
has a real $Re E_L(\vec r_k)$ and an imaginary $Im E_L(\vec r_k)$ parts.)

9) It are repeated steps 1)-8) an integer number of time steps
$\tau$ of integration of diffusion equation.
After that it are determined the mean values of quantities
$Re \overline E$, $Im \overline E$ and others on this integer number too. It is
redetermined the new value of $E_T$ in accordance with
 $(E_T)_{new}=[(E_T)_{old}+Re \overline E]/2$ accordingly with assumption that
 $Im \overline E <<Re \overline E$.
An integer number of time steps $\tau$ represents an one time block
 $\Delta t$.

10) Every time block  $\Delta t$ has the $N_c$ of initial configurations.
The $N_c$ of initial configurations are filled randomly by configurations of
just ended time block. The random choice of configurations consists of two
steps:\\
a) a random choice of a number of time step $\tau$ in every time
block $\Delta t$;\\
b) a random choice of configuration from $N$ configurations
at this fixed time step $\tau$.\\
 In this manner filled list of $N_c$
configurations will be initial one for the next time block $\Delta t$.

11) The repeating of big number of time blocks $\Delta t$
decreases essentially a correlation between configurations in neighbor
time blocks and provides a right calculation of mean quantities.

Let us discuss in detail the calculation of the mean quantities.
In general case (see \ci{4,7}) the mean value of the some quantity
$F(\vec R(t))$ with the complex distribution function $f(\vec R,t)$ at the
time $t$ of the running process is
\be
<F(t)>=\dsf{\ds\sum_{i=1}^M\exp[i\alpha(\vec R_i(t)]F(\vec R_i(t))}{
\ds\sum_{i=1}^M\exp[i\alpha(\vec R_i(t)]}.
\lab{21}
\ee
Here $\vec R_i(t)$ is the coordinates $\vec r_1,\vec r_2,...,\vec r_N$,
$N$ - number of particles of system,
 $M$ - number of configurations at time point $t$.
The particles coordinates $\vec r_1,\vec r_2,...,\vec r_N$
in (21) are weighted with probability $|f(\vec R_i,t)|$,
and the quantity $\alpha(\vec R_i(t))$ is a phase of the distribution function
$f(\vec R_i,t)$.

As it is clear from the integral expression \re{15}, the phase of
the distribution function at consequent time moment determines through the
phase of the Green function and the one of the distribution function at
previous time moment. So, we have:
\be
\alpha(\vec R_i(t+\tau))=\alpha_G(\vec
R_i(t+\tau),\vec R_j(t))+\alpha(\vec R_j(t)).
\lab{22}
\ee
Here  $\alpha_G(\vec
R_i(t+\tau),\vec R_j(t))$ is the phase of the Green function \re{14}
(with account of \re{16}),
and index  $j$ shows that $\vec R_j$ taken from
configurations at time moment $t$. By substituting \re{22} into \re{21}
we  have
\bea\ba{c} <F(t+\tau)>=\dsf{\ds\sum_{i=1}^M e^{i\alpha_G(\vec
 R_i(t+\tau), \vec R_j(t))+ i\alpha(\vec R_j(t))}F(\vec R_i(t+\tau))}{
\ds\sum_{i=1}^Me^{i\alpha_G(\vec R_i(t+\tau), \vec R_j(t))+
i\alpha(\vec R_j(t))}}=\\
=\dsf{\ds\sum_{i=1}^Me^{i\alpha_G(\vec R_i(t+\tau), \vec R_j(t))}F(\vec R_i(t+\tau))}{
\ds\sum_{i=1}^Me^{i\alpha_G(\vec R_i(t+\tau), \vec R_j(t))}}.
\lab{23}
\ea\eea
The mean quantity $<F(t+\tau)>$ in \re{23}
is determined only by the phase
 $\alpha_G$ of the Green function, because all quantities under sum
in the numerator and the denominator are weighted with the probability
$|f(\vec R_i(t+\tau),t+\tau)|$, i.e. at time moment
$t+\tau$ with new configurations  $\vec R_i(t+\tau)$.

Next, let us consider the expression for the phase of the Green function
\re{16}.
We have
\be
\alpha_G=\vec A_Q(\vec r,\vec r\ ')
(\vec r-\vec r\ '-D\tau Re \vec F_Q(\vec r\ '))-\tau Im E_L(\vec r).
\lab{24}
\ee
From \re{6} and \re{7} it is seen that the expressions $Re  \vec F_Q(\vec r\
'))$ and $E_L(\vec r)$ have no $\tau$ dependence.
From \re{17} one can show that
at  $\tau\ra 0$   $|\vec r|\ra |\vec r\ '|$ as $\tau^{1/2}$, because
$\chi$ has  $\tau^{1/2}$ dependence.
At the same limit
$\vec A_Q(\vec r,\vec r\ ')\ra \vec A_Q(\vec r\ ',\vec r\ ')$
(see the expression \re{13} for the $\vec A_Q(\vec r,\vec r\ ')$), i.e. there
is no $\tau$ dependence, also.
So, we have  $\alpha_G\sim\tau^{1/2}$.

\section{Analytical expressions for the algorithm quantities}

We take  trial wave function of electron
in the form:
 \be \Psi_T^*(\vec r)=\dsf{C}{a_H}\left(\dsf{\alpha
x+i\beta y}{a_H}\right)^m \exp\left(-\gamma\dsf{(x^2+ y^2)}{4a_H}\right).
\lab{26}
\ee
Here $C$ is normalization constant; $a_H=(\hbar/Mw_H)^{1/2}$
is magnetic length, where $w_H=|e|H/Mc$
- cyclotron frequency; $x$ and  $y$ the two spatial components of electron's
coordinate;
$m$ is orbital quantum number; $\alpha$, $\beta$ and $\gamma$
are arbitrary numerical constants.

At $\alpha=\beta=\gamma=1$ \re{26} gives an exact analytical expression of
electron ground state wave function. We deform
exact electron wave function, i.e. take as the input
$\alpha$, $\beta$ and $\gamma$ not equal to $1$, and then study a
relaxation of the trial function to the exact one via simulation process.

As length unit  we take $a_H$,
 as energy unit - $\hbar w_H/2$, and as time unit - $2/\hbar w_H.$

By substituting of the wave function \re{26} into \re{9}, we have
\bea\ba{l}
Re \vec F_Q(\vec r)=\dsf{1}{\alpha^2 x^2+\beta^2y^2}[
\vec i(2m\alpha^2x-\gamma \alpha^2x^3-\gamma\beta^2 xy^2)+\\
+\vec j(2m\beta^2y-\gamma \alpha^2yx^2-\gamma\beta^2 y^3)],\\
Im \vec F_Q(\vec r)=\dsf{2m\alpha\beta}{\alpha^2 x^2+\beta^2y^2}[
-y\vec i+x\vec j].
\lab{27}\ea\eea
Here and below $\vec i$ and $\vec j$
the unit vectors in $x$ and $y$ directions correspondingly.

For the  $\vec A_Q(\vec r,\vec r\ ')$  \re{13} gives
\bea\ba{l}
\vec A_Q(\vec r,\vec r\ ')=
\dsf{1}{\alpha^2 x'^2+\beta^2y'^2}[
\vec i(-2m\alpha\beta y'+y(\alpha^2x'^2+\beta^2 y'^2))+\\
+\vec j(2m\alpha\beta x'-x(\alpha^2x'^2+\beta^2 y'^2))].
\lab{28}\ea\eea
The wave function \re{26} and eq. \re{7}
(here $V(r)=0$) gives
\be
Re E_L(\vec r)=Re E_L^{(1)}(\vec r)+Re E_L^{(2)}(\vec r),
\lab{29}\ee
where
$$
Re E_L^{(1)}(\vec r)=\gamma(m+1)+(x^2+y^2)\left[\dsf{1-\gamma^2}{4}-
\dsf{m\alpha\beta}{\alpha x^2+\beta^2y^2}\right],
$$
$$
Re E_L^{(2)}(\vec r)=\dsf{m(m-1)(\beta^2-\alpha^2)
(\alpha x^2-\beta^2y^2)}{(\alpha^2 x^2+\beta^2y^2)^2};
$$
and also
\be
Im E_L(\vec r)=\dsf{mxy(\beta^2-\alpha^2)}{\alpha^2 x^2+\beta^2y^2}\left[
1-\dsf{(m-1)2\alpha\beta}{\alpha^2 x^2+\beta^2y^2}\right].
\lab{30}\ee

By using formulas \re{27}-\re{30}, it is easy to write the analytical
expressions for other quantities.

\section{The discussion of the simulation results.}

The simulation of the ground state of one electron in magnetic
field is a simple problem
and  was performed
on the usual personal computer.

Figs.1-4 are presented the result of the calculations of
the real $Re \overline E$ and imaginary part $Im \overline E$
of the ground state energy
for different orbital quantum numbers $m$ and different starting
trial functions defined by the parameters $\alpha , \beta , \gamma$
as a functions of the
number $T$ of time blocks $\Delta t$ (see point 9) of algorithm)
( a full number of time blocks are chosen
ten in each running).

In every time block $\Delta t$
initial number of configurations $N_c$
was chosen equal 1000 and number of time steps $\tau$
equal 100. We everywhere suggested $\tau$
equal 0.01.
Typical mean number of the population number  $N$ ( see point 6) of the algorithm)
was around 2500 at initial blocks $\Delta t$ and around 1000 at final
blocks $\Delta t$.

Fig.1 present $Re \overline E$   and Fig.2 present $1000 Im \overline E$
for $m=0,4,8$ with parameters
$\alpha = \beta =\gamma =1$  as a functions of $T.$

It is seen from Figs.1,2  $Re \overline E$ is approaching fast to
the exact value of ground state energy and
reproduce very well an orbital quantum number $m$ degeneracy,
while the imaginary part of energy
$Im \overline E \sim 10^{-3}Re \overline E.$

The additional calculations shows, that the increasing
of the statistics, i.e. the increasing of  $N_c$ and a number of
steps  $\tau$ in each block $\Delta t$, decreases
$Im \overline E$ more.

The Figs.3,4 represent the $Re \overline E$ and $Im \overline E$
for the different choices of the $m=0$ trial function.
From these Figs. we can
see that the deformation of  wave function in some limits does not
alter of the relaxation of the running process to exact final state.

Fig.5 present
the initial uniform spatial distribution of electron before simulation,
while Fig.6 -
the final spatial distribution of electron after simulation for $m=13$.

In this simulation the trial wave function  $\Psi_T^*$ is taken
with parameters $\alpha = \beta =\gamma =1$, $m=13$ and  $N_c=500$.
Since the trial function coincide with exact wave function we have to
expect fast convergence of initial distribution to exact one if the
algorithm is correct.
In accordance with analytical solution (see \ci{9}) the  spatial
distribution for one electron in this case must have a ring like form with
mean radius  $(2m+1)^{1/2}$ and wide 1 ( in magnetic length
$a_H$ units).
As it is seen from Fig.6, there is an excellent agreement between
simulation and analytical results.

We suppose also further tests of the method
by simulations of anyons \ci{AMN2}.

\begin{figure}
\begin{center}
\parbox{10cm}{\epsfxsize=9.cm \epsfysize=9.cm \epsfbox[5 5 500 500]
{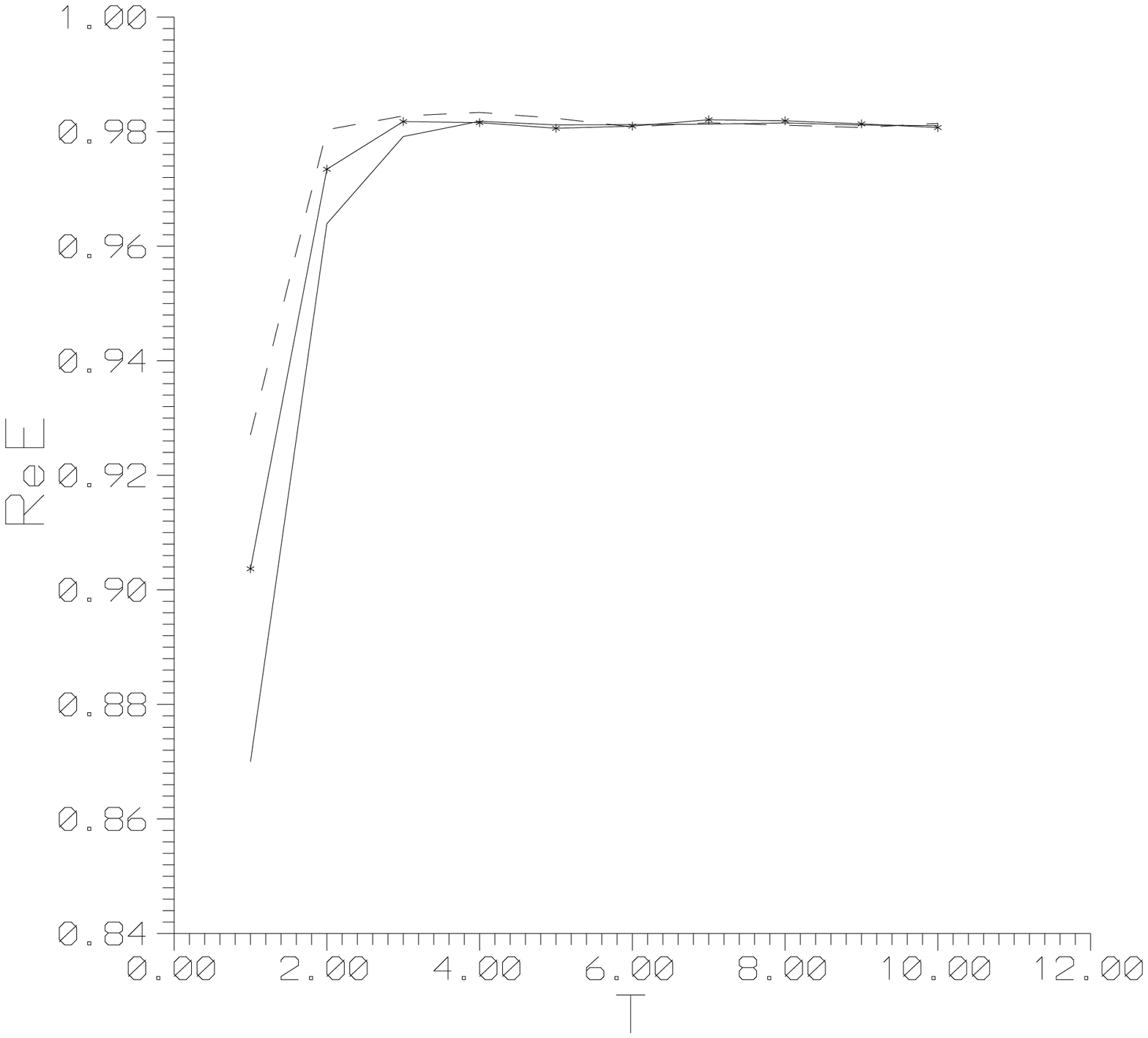}{}}
\end{center}
\caption{
The real part
of the ground state energy $Re \overline E$
as a functions of the
number $T$ of time blocks $\Delta t$ (see point 9) of algorithm)
for different orbital quantum numbers $m$.
Solid line - $m=0$, marked solid line - $m=4$, dashed line - $m=8$.
}
\end{figure}
\begin{figure}
\begin{center}
\parbox{10cm}{\epsfxsize=9.cm \epsfysize=9.cm \epsfbox[5 5 500 500]
{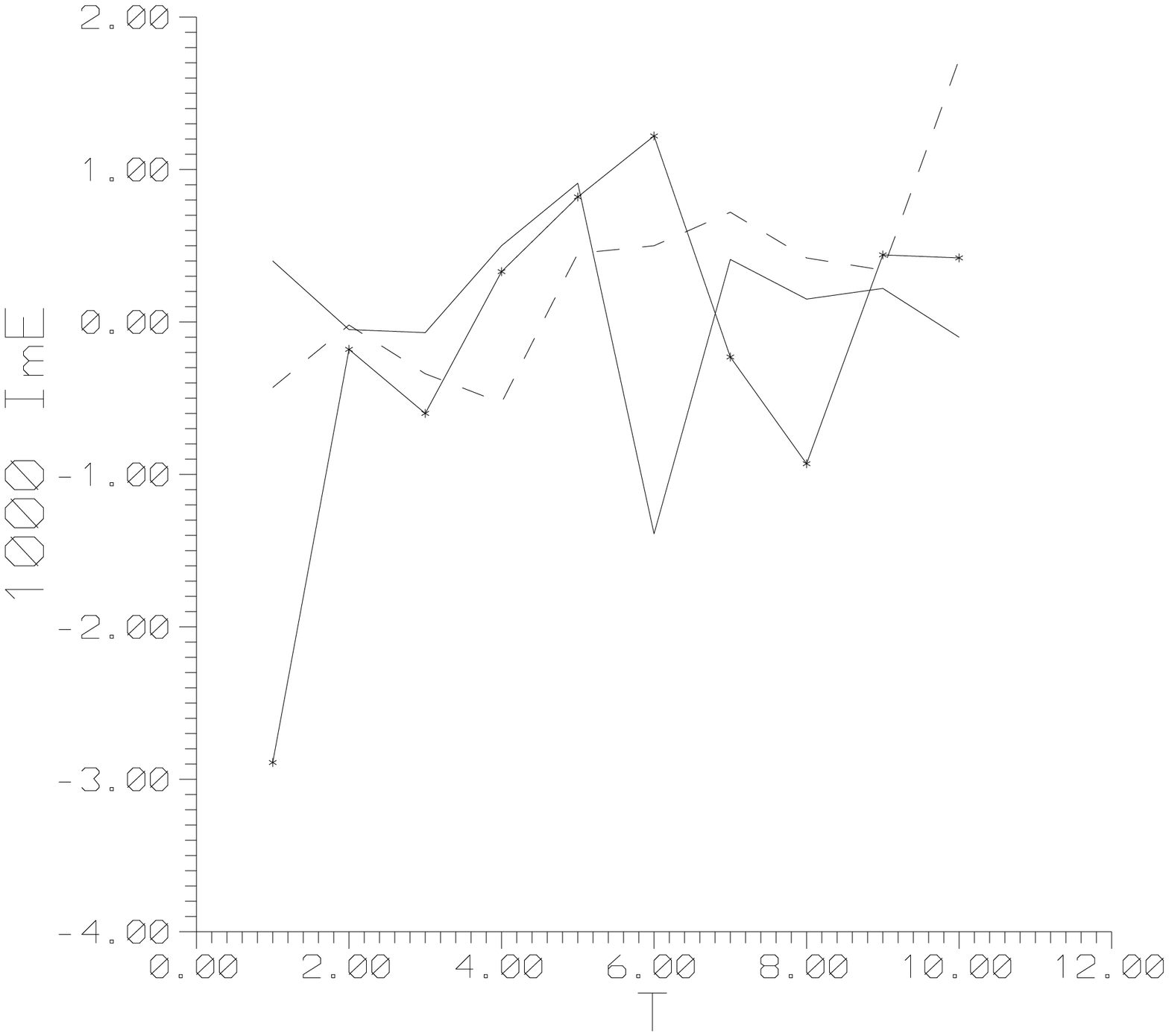}{}}
\end{center}
\caption{
The same as Fig.1 but for
$1000 Im \overline E$.}
\end{figure}
\begin{figure}
\begin{center}
\parbox{10cm}{\epsfxsize=9.cm \epsfysize=9.cm \epsfbox[5 5 500 500]
{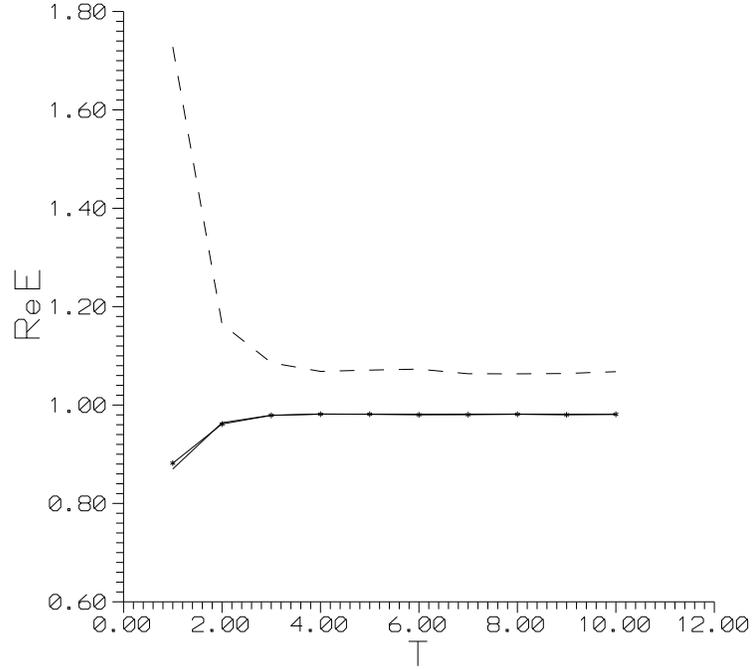}{}}
\end{center}
\caption{
The real part
of the ground state energy $Re \overline E$
for
different starting
parameters $\alpha , \beta , \gamma$ and the same $m=0$
as a functions of the
number $T$ of time blocks $\Delta t$.
Solid line - $\alpha = \beta =\gamma = 1$,
marked solid line - $\alpha =1.4, \beta =\gamma = 1$,
dashed line - $\alpha = \beta =1, \gamma = 0.9$.}
\end{figure}
\begin{figure}
\begin{center}
\parbox{10cm}{\epsfxsize=9.cm \epsfysize=9.cm \epsfbox[5 5 500 500]
{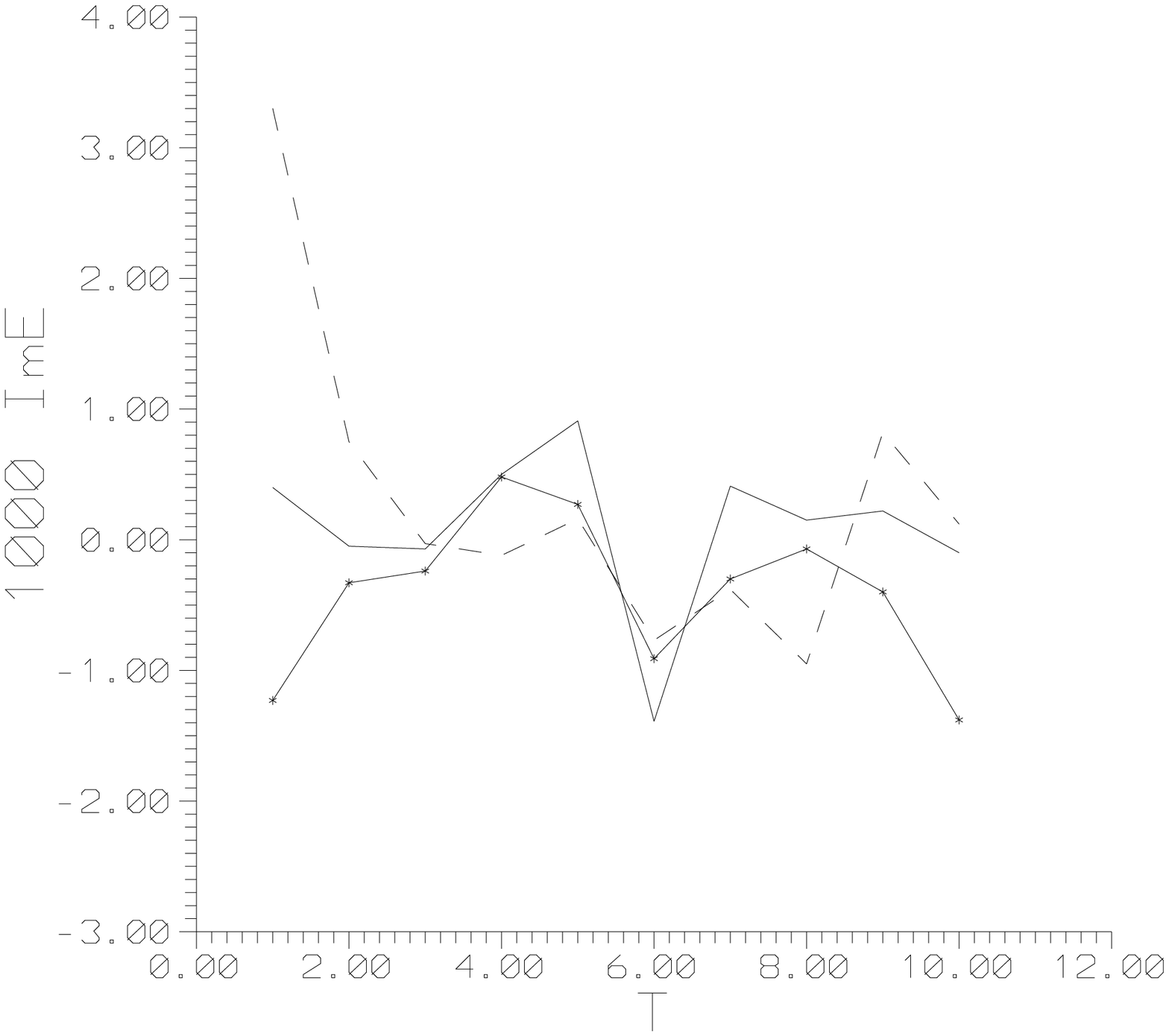}{}}
\end{center}
\caption{
The same as Fig.3 but for
$1000 Im \overline E$.}
\end{figure}

\begin{figure}
\begin{center}
\parbox{10cm}{\epsfxsize=10.cm \epsfysize=10.cm \epsfbox[5 5 500 500]
{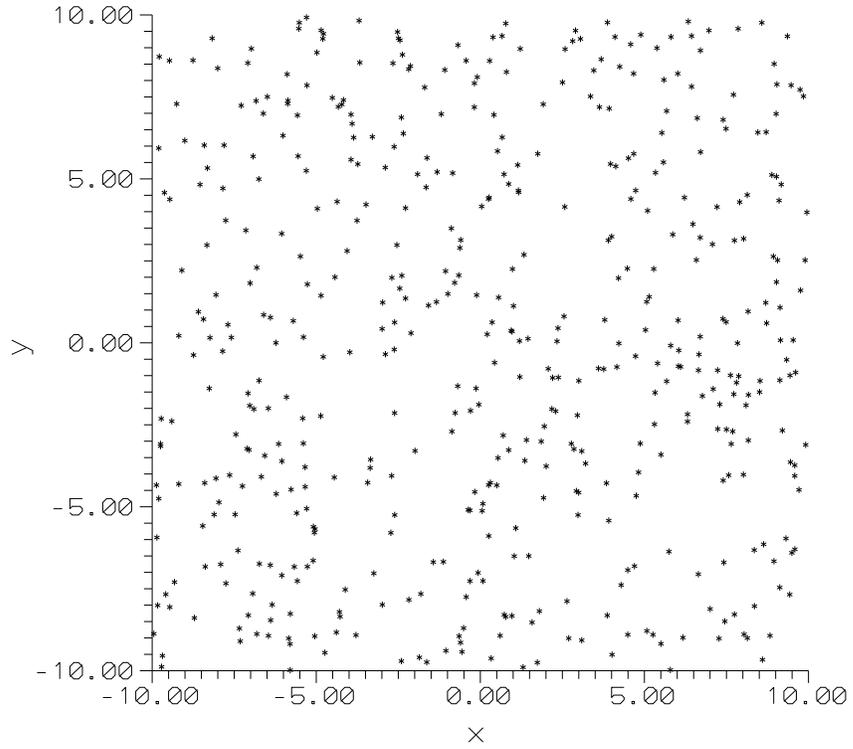}{}}
\end{center}
\caption{Initial spatial distribution of the electron.}
\end{figure}
\begin{figure}
\begin{center}
\parbox{10cm}{\epsfxsize=10.cm \epsfysize=10.cm \epsfbox[5 5 500 500]
{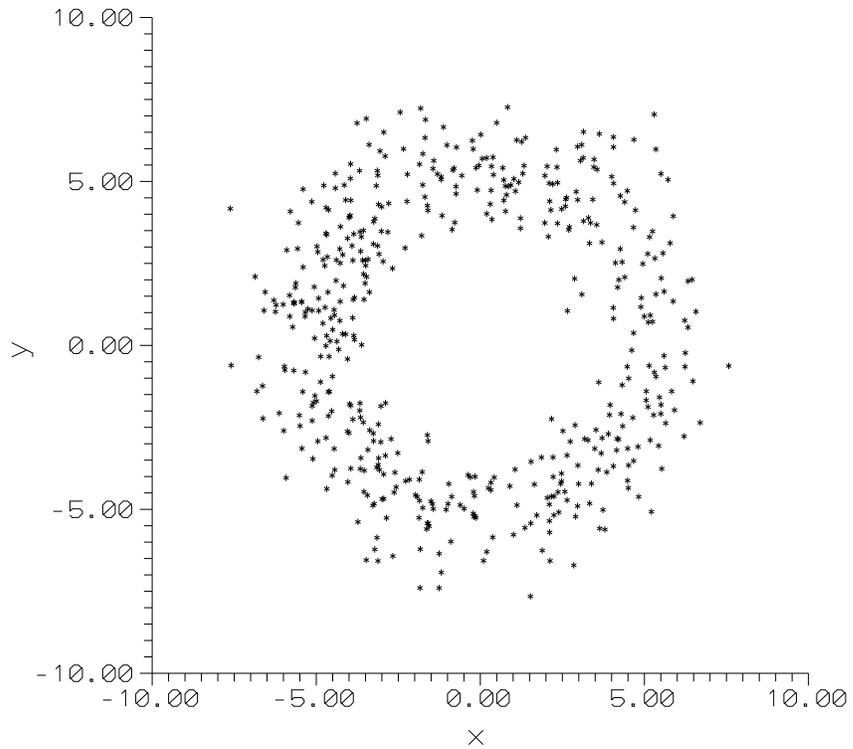}{}}
\end{center}
\caption{The result of the simulation:
final spatial distribution of
the electron in uniform magnetic field  at $m=13$. The  length
unity is $a_H$, initial number of points $N_c=500$.}
\end{figure}

\edc